\documentclass[aps,reprint,superscriptaddress,pre,nofootinbib,citeautoscript,oneside]{revtex4-1}
\usepackage[utf8]{inputenc}
\usepackage{mathtools}

\usepackage{graphicx}
\usepackage{dcolumn}
\usepackage{bm}
\usepackage{xcolor}
\usepackage{amsmath,amssymb,amsthm}
\usepackage{hyperref}

\begin{document}

\title{Comparing the efficiency of forward and backward contact tracing}

\date{\today}

\author{Jonas L. Juul}
\email{jjuul@cornell.edu}
\affiliation{Center for Applied Mathematics, Cornell University, Ithaca, New York 14853, USA}
\author{Steven H. Strogatz}
\affiliation{Center for Applied Mathematics, Cornell University, Ithaca, New York 14853, USA}
\newcommand{\ER}{Erd\H{o}s--R\' enyi{}}
\newcommand{\BA}{Barab{\'a}si--Albert{}}

\begin{abstract}
Tracing potentially infected contacts of confirmed cases is important when fighting outbreaks of many infectious diseases. 
The COVID-19 pandemic has motivated researchers to  examine how different contact tracing strategies compare in terms of effectiveness (ability to mitigate infections) and cost efficiency (number of prevented infections per isolation). 
Two important strategies are so-called forward contact tracing (tracing to whom disease spreads) and backward contact tracing (tracing from whom disease spreads). 
Recently, Kojaku and colleagues reported that backward contact tracing was ``profoundly more effective'' than forward contact tracing, that contact tracing effectiveness ``hinges on reaching the `source' of infection'', and that contact tracing outperformed case isolation in terms of cost efficiency.
Here we show that these conclusions are not true in general. They were based in part on simulations that vastly overestimated the effectiveness and efficiency of contact tracing. 
Our results show that the efficiency of contact tracing strategies is highly contextual; faced with a disease outbreak, the disease dynamics determine whether tracing infection sources or new cases is more impactful. 
Our results also demonstrate the importance of simulating disease spread and mitigation measures in parallel rather than sequentially. 
\end{abstract}
\maketitle
\section{Introduction}
To combat the spread of an infectious disease, contact tracing can be a useful mitigation tool. Contact tracing aims to identify infectious individuals through their social contacts; when an infected person has been identified, their close social contacts are traced and asked to test and possibly get treatment or quarantine. By proactively tracing and testing potentially infected people, public health officials can reduce the transmission of the infectious disease. For diseases such as SARS-CoV-2, where many transmissions take place before the infectious person develops symptoms~\cite{he_temporal_2020,oran_prevalence_2020,arons_presymptomatic_2020}, contact tracing is critical in reducing further spread~\cite{moghadas_implications_2020,grassly_comparison_2020,noauthor_report_2020}.

During the last two decades, mathematical epidemiologists have sought to understand in what circumstances contact tracing is effective~\cite{muller_contact_2021}. Among other things, researchers have investigated how the ability of contact tracing to curb the spread of disease is influenced by the network structure of social interactions~\cite{huerta_contact_2002,eames_modeling_2002,eames_contact_2003,kiss_disease_2005,kiss_infectious_2006,eames_contact_2007,kiss_effect_2008,house_impact_2010,okolie_exact_2020}, disease characteristics such as infectiousness profiles or case report statistics~\cite{klinkenberg_effectiveness_2006,browne_modeling_2015,peak_comparing_2017,baumgarten_epidemics_2021,juul_are_2021}, different choices in how contact contact tracing is carried out~\cite{eames_contact_2007,kiss_effect_2008,meister_optimizing_2021}, and specific outbreak scenarios (real or imagined)~\cite{kaplan_emergency_2002,eames_assessing_2010,swanson_contact_2018}. The findings of  these studies have led to both concrete recommendations in case of bioterrorism~\cite{kaplan_emergency_2002}, and more qualitative insights such as that contact tracing effectiveness can be affected by clustering and heterogeneity in population contact network structures~\cite{eames_contact_2007,house_impact_2010}.

The COVID-19 pandemic spurred a range of theoretical epidemiological investigations. Topics included vaccination prioritization strategies~\cite{bubar_model-informed_2021}, the importance of accuracy and turnaround time in diagnostic tests~\cite{mina_rethinking_2020,larremore_test_2021,juul_are_2021}, how standard confidence intervals can obscure extremes of epidemic projections (and what the alternatives are when presenting such projections to decision makers)~\cite{juul_fixed-time_2021}, how schools and other institutions can be reopened safely following lockdowns~\cite{karin_cyclic_2020,mcgee_model-driven_2021,frazier_modeling_2022}, and how superspreading influences disease transmission and control measures~\cite{nielsen_covid-19_2021, sneppen_overdispersion_2021, althouse_superspreading_2020}. Attention was also paid to contact tracing and its effectiveness. Arguably, the most influential theoretical contact tracing study during the pandemic was that authored by Kojaku \textit{et al.}~\cite{kojaku_effectiveness_2021}.

Kojaku \textit{et al.}~\cite{kojaku_effectiveness_2021} investigated what makes contact tracing efficient in networked populations. They showed that backward contact tracing preferentially leads to high-degree infected individuals \--- superspreaders \--- with a sampling bias stronger than the celebrated friendship paradox~\cite{kojaku_effectiveness_2021}. The friendship paradox states that ``your friends have on average more friends than you do''~\cite{feld_why_1991}; the reason being that (not considering possible degree correlations) following a random edge from a node leads to a degree-$k$ node with probability proportional to $kp_k$, where $p_k$ is the probability that a uniformly random node has degree $k$. The statistical arguments presented by Kojaku \textit{et al.}~\cite{kojaku_effectiveness_2021} show that tracing backward in the transmission tree (a rooted, directed tree illustrating who infected whom) leads to degree $k$-nodes with probability proportional to $k^2p_k$ \--- a much stronger bias than that underlying the friendship paradox. This important result formalizes how effective backward tracing is at uncovering superspreaders.

In addition to the derivation of the backward-tracing sampling bias, Kojaku \textit{et al.}~\cite{kojaku_effectiveness_2021} also used numerical simulations to make general conclusions about the effectiveness of different disease mitigation strategies. One of their main findings was that ``compared to `forward' contact tracing (tracing to whom disease spreads), `backward' contact tracing (tracing from whom disease spreads) is profoundly more effective.'' Indeed, this quote was the center of most coverage of the study in news outlets and on social media~\cite{noauthor_altmetric_nodate}.
They also reported that the efficiency of contact tracing ``hinges on reaching the source of infection'' and that contact tracing beat case isolation in terms of cost efficiency. (Here, case isolation means quarantining confirmed cases without proactively using contact tracing to discover new cases through those already found). 

In this paper, we demonstrate that Kojaku \textit{et al.}'s conclusions about the superiority of backward tracing over forward tracing and case isolation  are not correct in general. Their  conclusions rely on simulations that systematically overestimate the efficiency and effectiveness of contact tracing. 
The overestimate stems mainly from the \emph{sequential} nature of how Kojaku \textit{et al.}~\cite{kojaku_effectiveness_2021} implement contact tracing and case isolation in their simulations: they first run their model epidemic to completion; only after that do they perform the model versions of case isolation and contact tracing. But in reality, these two processes occur contemporaneously rather than sequentially.  

More precisely, Kojaku \textit{et al.}~\cite{kojaku_effectiveness_2021} simulate epidemics involving four classes of people relative to the disease, namely those who are susceptible (S), exposed (E), infected (I), and recovered or removed (R). They begin by simulating unconstrained SEIR epidemics spreading on various network topologies. 
From these simulations, they then obtain transmission trees.  
Finally, contact tracing and case isolation are simulated using these transmission trees: If an infectious node is successfully identified and isolated, Kojaku \textit{et al.}~\cite{kojaku_effectiveness_2021} assume that all the nodes downstream in the transmission tree (the `descendants' that would have otherwise been infected) are thereby precluded from getting infected.  
But this assumption ignores the fact that a downstream node may well have other infected neighbors, each of which could potentially pass the infection on to it, even if the traced ancestor has been isolated. 

The overestimation of contact tracing efficiency and effectiveness is likely to be large because descendant distributions are heavy-tailed~\cite{juul_descendant_2020}. 
In reality, the epidemic unfolds side-by-side with mitigation interventions like contact tracing. 
For this reason, simulating disease mitigation and epidemic spreading in parallel would provide a more accurate estimate of the efficacy of contact tracing. 

In this paper, we examine contact tracing efficiency by simulating disease spread and mitigation measures in parallel rather than sequentially. This choice reduces the estimated efficiency of contact tracing by an order of magnitude compared to Kojaku \textit{et al.}'s estimates~\cite{kojaku_effectiveness_2021}, and demonstrates that backward contact tracing can be less efficient than forward contact tracing. Correcting another shortcoming of Kojaku \textit{et al.}'s simulations---that they never release susceptible nodes that were traced and quarantined---further decreases contact tracing efficiency and demonstrates that case isolation can be more efficient than contact tracing.

\section{Simulating epidemics and mitigation strategies}
We simulate epidemics unfolding on a {\BA} network~\cite{barabasi_emergence_1999} like the one used by Kojaku \textit{et al.}~\cite{kojaku_effectiveness_2021}. The network has $250\,000$ nodes with average degree $4$.
Our simulation progresses in discrete time steps. 
We assume that each infected node first spends $t_E$ days as presymptomatic and then is infectious for $t_I$ days. Each of these times is drawn from probability distributions of our choice. Unless otherwise stated, we draw $t_E$ from an exponential distribution with mean $4$ days and $t_I$ from a lognormal distribution recently reported to resemble the incubation period of COVID-19~\cite{mcaloon_incubation_2020}. We note that our results are robust to changes in these distributions: excluding the exposed compartment entirely or drawing $t_E$ and $t_I$ from the same exponential distribution with mean $4$ days yields similar results. 

In our model, we assume that an infectious node infects each susceptible neighbor with probability $p_I = q(t-t_0)\bar{p_I}$ at each time step. 
Here $\bar{p_I}$ is an average infectivity and $q(t-t_0)$ is a normalized function that expresses the relative likelihood of the node infecting a neighbor each day after time step $t_0$ when the node itself became infectious. We choose $\bar{p_I} = \frac{1.0}{(\bar{k} -1)}$, where $\bar{k}$ is the mean degree of the network. 

In addition to the disease dynamics described above, we also simulate disease mitigation measures. Like Kojaku \textit{et al.}~\cite{kojaku_effectiveness_2021}, we assume that an infected node is identified with probability $p_s$ at symptom onset and that each of its neighbors (infectious or not) is successfully traced with probability $p_t$. When a node is traced, we add it to a contact list with some weight $w$.  To facilitate simulation of various contact tracing strategies, we treat $w$ as adjustable.  If $w=1$ for all nodes, the contact tracing scheme reduces to that of Kojaku \textit{et al.}~\cite{kojaku_effectiveness_2021}; other choices of $w$ allow us to implement backward contact tracing and forward contact tracing, as we discuss below.

We compare the impact of backward contact tracing and forward contact tracing by carrying out two kinds of simulations. In one, we simulate backward contact tracing by allowing only the direct source of infection to be traced ($w=w_{\rm parent}=1$ if a traced node is the direct source of infection, and $w=1-w_{\rm parent}=0$ otherwise). In the other, we simulate forward contact tracing by allowing any neighbor other than the direct source of infection to be traced ($w=w_{\rm parent}=0$ if a traced node is the direct source of infection, and $w=1-w_{\rm parent}=1$ otherwise). 
At each time step, after the above-described contact tracing has been carried out, we sum up the weights that each node is listed with in the contact list. 
Like Kojaku \textit{et al.}~\cite{kojaku_effectiveness_2021}, we then isolate the $n=30$ nodes with the highest sum of weights in that list.  We clear the contact list at the beginning of each time step. 


\section{Results}

We now present results indicating that (i) backward contact tracing is not necessarily more effective or efficient than forward contact tracing and (ii) contact tracing is not necessarily more efficient than case isolation. 
Our simulations also indicate that Kojaku \textit{et al.}'s simulation choices cause their estimate of contact tracing efficiency to be an order of magnitude too high.
\subsection{Comparing the efficiency of backward and forward tracing}
To compare backward contact tracing to forward contact tracing, we simulate the mitigation of disease outbreaks with each of these contact tracing strategies. Backward contact tracing is simulated by allowing only the direct source of infection to be traced from an infected node; forward contact tracing is simulated by allowing every neighbor except the direct source of infection to be traced from an infected node. In practice, we implement backward contact tracing by weighting a parent by $w_{\rm parent}=1$ in the contact list if it is traced through one of its children, and weighting all non-parents by $w_{\rm non-parent}=1-w_{\rm parent}=0$.

\begin{figure*}
    \centering
    \includegraphics[width=0.48\linewidth]{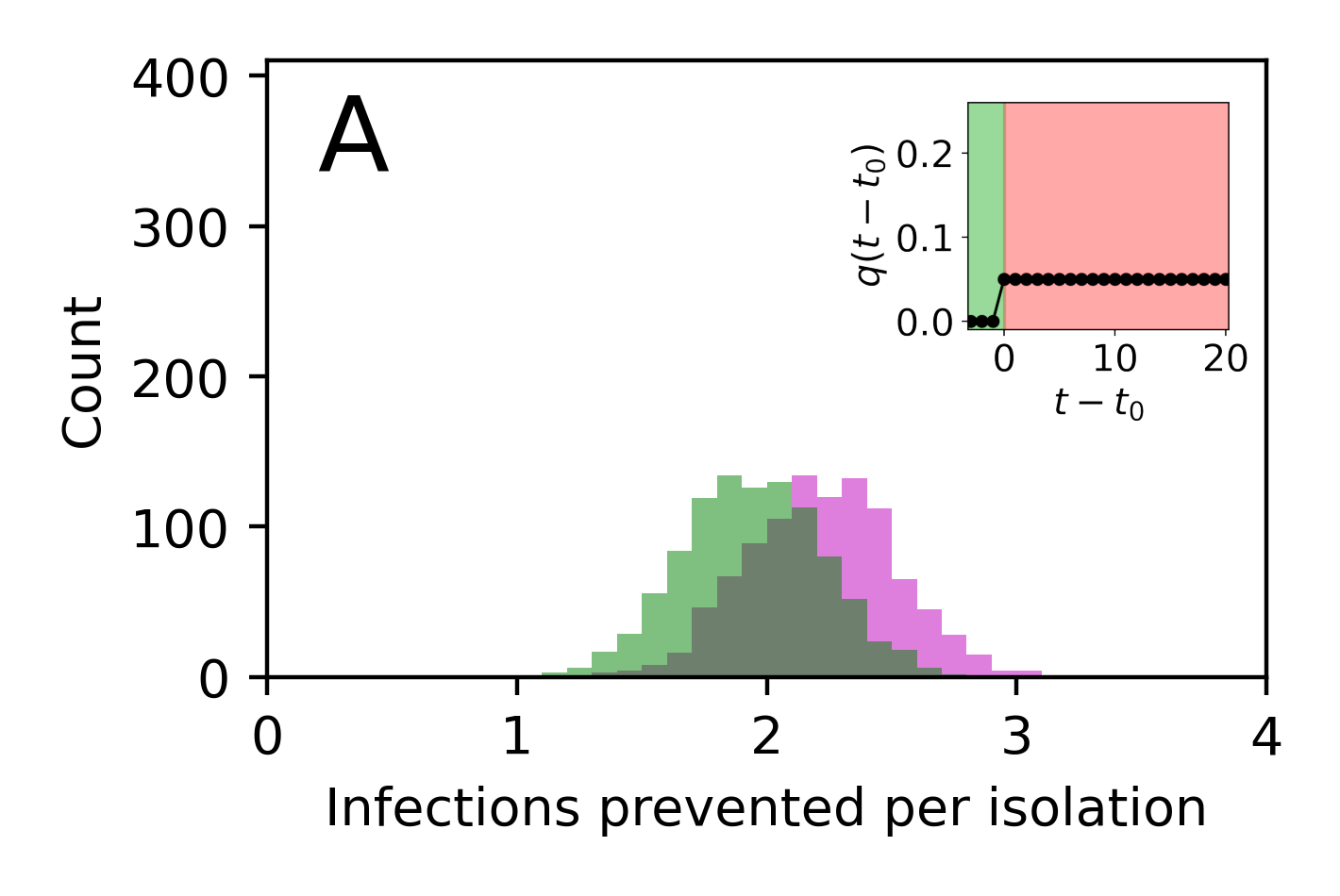}
    \hfill
    \includegraphics[width=0.48\linewidth]{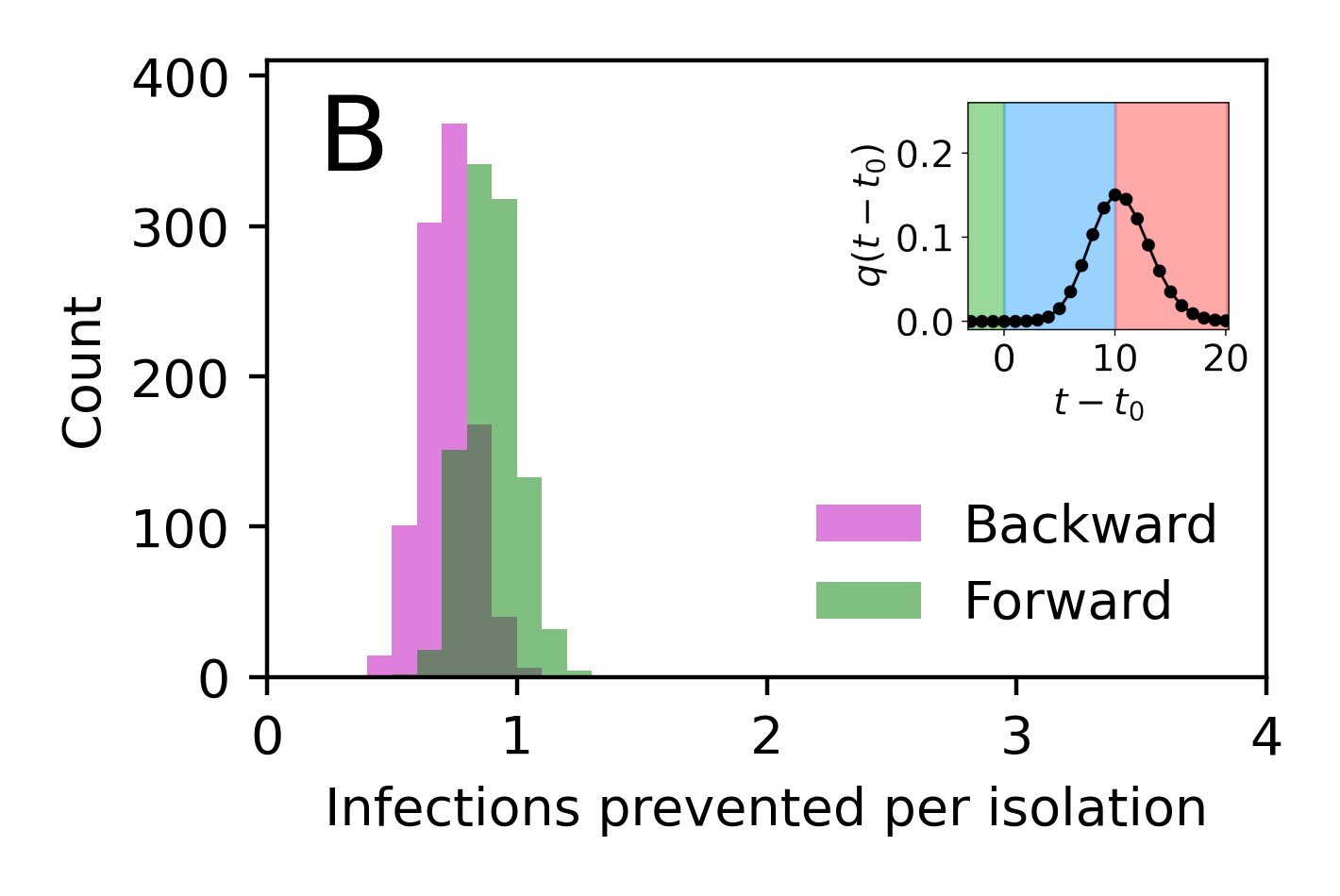}

    \caption{\textbf{Efficiency of contact tracing for two epidemic models with different patterns of infectiousness.} Figure insets illustrate the assumed infectiousness pattern. Vertical values illustrate infectiousness on day $t-t_0$ after the node became infectious, and inset background color illustrates the disease state of a node on that day (Green: Exposed; Blue: Asymptomatic and infectious; Red: Symptomatic and infectious). In the main panels of the figure, we plot the number of infections that were prevented by two types of contact tracing for each person that was isolated in a simulation. Magenta histograms show results for an idealized backward contact tracing scheme in which only the direct source of infection (the `parent' node) of an identified infected node  can be traced; in the green histograms, any neighbor except the direct source of the infection  can be traced. We take the number of prevented infections to be the difference in nodes that got exposed to the disease when simulating the epidemic with parameters $p_s=0$, $p_t=0$ and $p_s=0.05$, $p_t=0.50$ (and otherwise identical initial conditions). The histograms show values obtained for $1\,000$ different simulations.  \textbf{A} Constant infectiousness model. An infectious node is always symptomatic and infects each susceptible neighbor with equal probability on each of its infectious days. Backward contact tracing is the more efficient mitigation strategy, as shown by the relative positions of the magenta and green histograms. \textbf{B} Skewed infectiousness model. An infected node is asymptomatic during its first half of its infectious period and its infectiousness peaks around symptom onset. Backward contact tracing is the less efficient strategy here. }
    \label{fig:flat_vs_empirical}
\end{figure*}

For the simulation of each contact tracing strategy, we use the parameters $p_s = 0.05$ and $p_t=0.50$, and simulate two different epidemic models: a `constant infectiousness' model and a `skewed infectiousness' model. The constant infectiousness model assumes that each infected node has a constant probability of passing the infection along to each of its susceptible neighbors during the days when the node is infectious and symptomatic (Fig.~\ref{fig:flat_vs_empirical}A). In the skewed infectiousness model, inspired by known properties of COVID-19, each infected node is asymptomatic during the first half of its infectious period, and its infectiousness peaks around symptom onset (Fig.~\ref{fig:flat_vs_empirical}B). Furthermore, for the skewed infectiousness model, we use an empirically estimated time-dependent transmissibility of COVID-19 for the function $q(t-t_0)$~\cite{he_temporal_2020,ashcroft_covid-19_2020}.

Figures~\ref{fig:flat_vs_empirical}A and B compare the efficiency of both kinds of contact tracing (with only parents or no parents) for the two infectiousness models. For the constant infectiousness model (Fig. ~\ref{fig:flat_vs_empirical}A), each isolation results in around $2$ prevented infections. For the skewed infectiousness model (Fig. ~\ref{fig:flat_vs_empirical}B), each isolation results in around 1 prevented infection. These numbers are an order of magnitude smaller than the approximately $20$ prevented infections per isolation estimated by Kojaku \textit{et al.}~\cite{kojaku_effectiveness_2021} for a constant infectiousness model.

Another point to notice is the relative positions of the magenta and green histograms in Figs.~\ref{fig:flat_vs_empirical}A and B. Backward contact tracing is the more efficient disease-mitigation strategy in the constant infectiousness model (Fig.~\ref{fig:flat_vs_empirical}A), but the less efficient strategy in the skewed infectiousness model (Fig.~\ref{fig:flat_vs_empirical}B). This reversal confirms the intuitive arguments made above. 

The observations presented above for the efficiency of the two contact tracing schemes can also be made for the effectiveness of the strategies. For the constant infectiousness model, the mean number of infected in a simulation is $94304 \pm 22$ for forward contact tracing and $93454 \pm 24$ for backward contact tracing. This makes backward contact tracing the more effective strategy, just like it was the more efficient strategy. For the skewed infectiousness model, however, the number of infected in a simulation changes to $92851 \pm 23$ for forward contact tracing and $94357 \pm 22$ for backward contact tracing, making forward contact tracing the more effective strategy in this case.

\subsection{Comparing case isolation and contact tracing}
Kojaku \textit{et al.} overestimate the efficiency of contact tracing by simulating disease spread and mitigation measures sequentially rather than in parallel. As we shall later demonstrate, this sequential implementation is what causes an overestimation of contact tracing efficiency by an order of magnitude. On top of that, another assumption incorrectly inflates Kojaku \textit{et al.}'s estimate of contact tracing efficiency. Correcting that assumption reveals that case isolation can be more cost efficient than contact tracing, contrary to Kojaku \textit{et al.}'s conclusion. 

\begin{figure}[]
    \centering
    \includegraphics[width=0.96\linewidth]{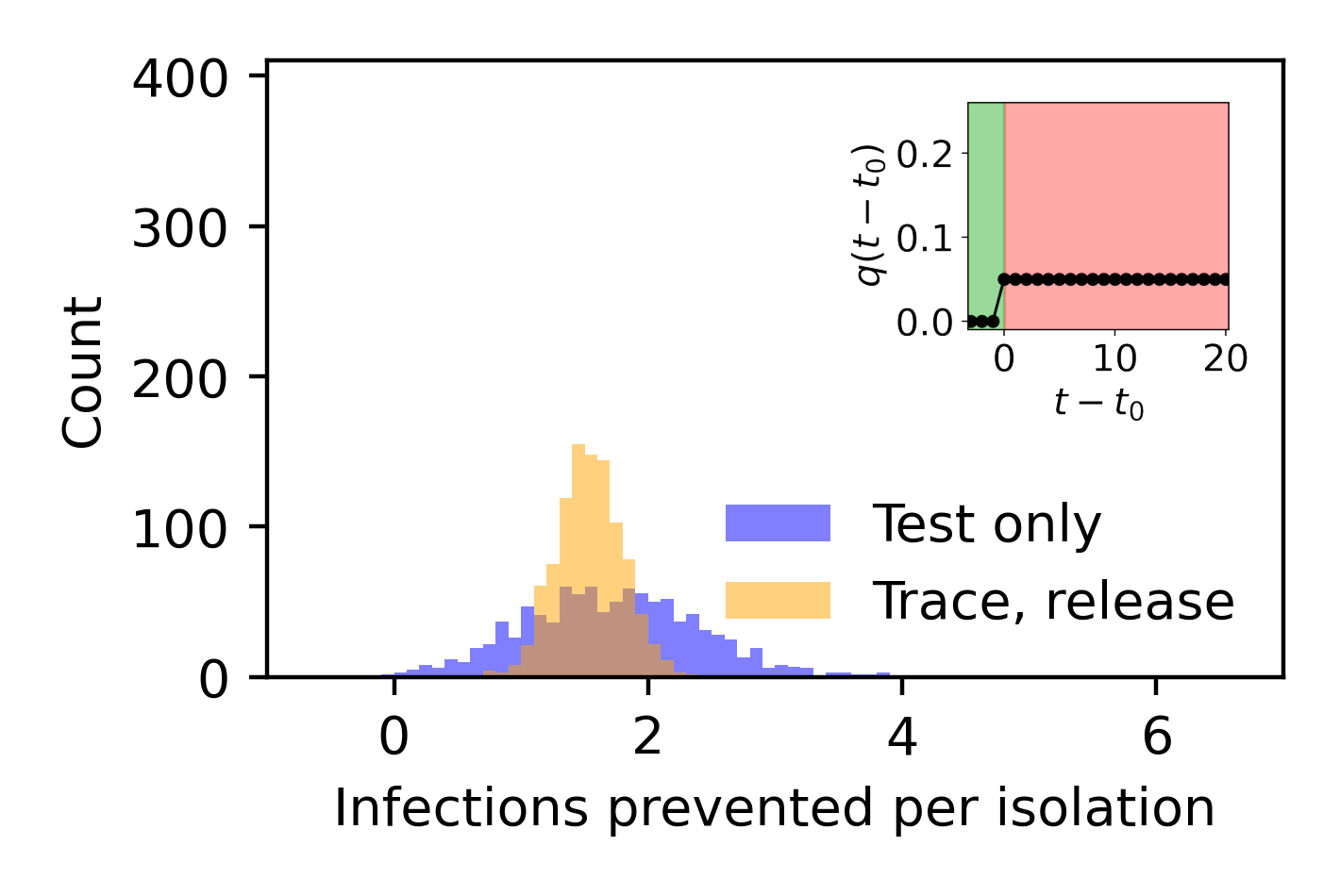}
    \caption{
    \textbf{Efficiency of case isolation vs.\ contact tracing when susceptible quarantined nodes are eventually released.} Figure inset illustrates the assumed infectiousness pattern. Vertical values illustrate infectiousness on day $t-t_0$ after the node became infectious, and inset background color illustrates the disease state of a node on that day (Green: Exposed; Red: Symptomatic and infectious). In the main panel of the figure, the yellow histogram plots the number of infections that were prevented by contact tracing for each person that was isolated in a simulation. We allow for both forward and backward contact tracing in this figure and release a traced and isolated node after $4$ days of isolation if it is not infectious. The blue histogram plots the number of infections that were prevented by case isolation for each person that was isolated in a simulation. We take the number of prevented infections to be the difference in nodes that got infected by the disease when simulating the epidemic with parameters $p_s=0$, $p_t=0$ and $p_s=0.05$, $p_t=x$ (where $x$ is $0.50$ for the yellow histogram and $0$ for the blue histogram) (and otherwise identical initial conditions). The histograms show values obtained for $1\, 000$ different simulations. In this case, the mean infections prevented per isolation is $1.72 \pm 0.02$ for case isolation (blue histogram) and $1.54\pm 0.01$ (yellow histogram), indicating that for these combinations of disease and mitigation measures, case isolation is more efficient than contact tracing.
}
    \label{fig:test_vs_tracerelease}
\end{figure}

A key unrealistic assumption in Kojaku \textit{et al.}'s~\cite{kojaku_effectiveness_2021} simulation of contact tracing and case isolation is that quarantined nodes remain in quarantine for all time \--- even if they are susceptible. This assumption, again, increases the estimated efficiency of contact tracing.
If we relax this unrealistic assumption, and instead assume that quarantined nodes are released after $4$ days if they are neither exposed nor infectious, case isolation becomes more efficient than contact tracing in terms of the number of prevented infections per isolation. This result is shown in Fig.~\ref{fig:test_vs_tracerelease}.
Thus, Kojaku \textit{et al.}~\cite{kojaku_effectiveness_2021} not only overestimated the numerical value of the estimates of contact tracing efficiency. They also overestimated the efficiency of contact tracing relative to other mitigation measures.

\begin{figure}
    \centering
    \includegraphics[width=0.96\linewidth]{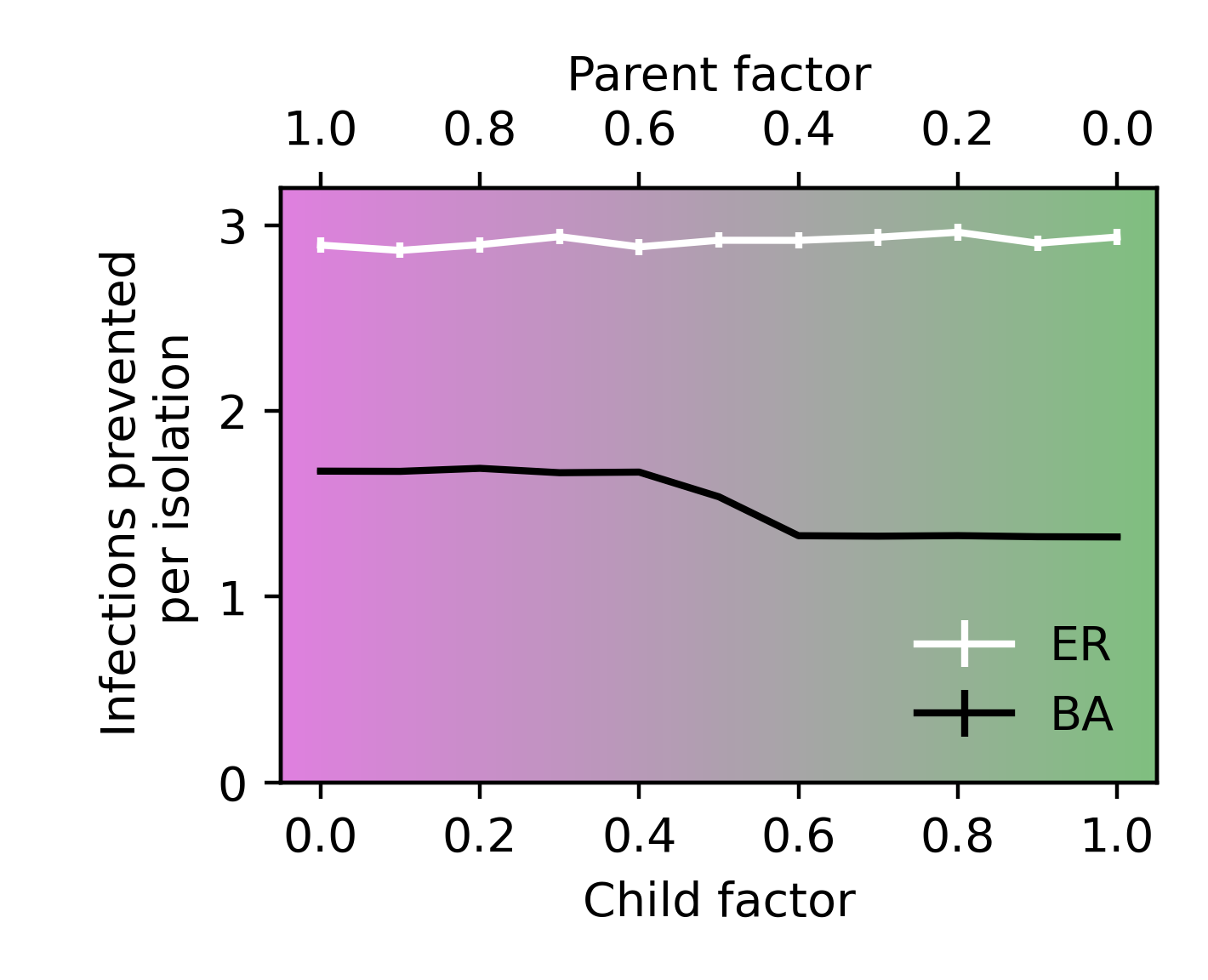}
    \caption{\textbf{Efficiency of contact tracing for combinations of backward and forward contact tracing.} We simulate a constant infectiousness model on two different networks with similar mean degree: a {\BA} network and and {\ER} network. With parameters $p_s=0.05$ and $p_t=0.50$, we simulate different combinations of forward and backward tracing. We combine backward and forward tracing by varying a parameter $w_{\rm parent}$ (``parent factor'' on the top horizontal axis). When a node is traced, we add it to the contact list with weight $w_{\rm parent}$ if it was the direct source of infection for the node it was traced through; otherwise, we add it to the contact list with weight $1-w_{\rm parent}$ (``Child factor'' on the bottom horizontal axis). After each time step, we add the weights of each node in the contact list. We then isolate the $30$ nodes with the highest sum of weights and clear the contact list before the next time step.  For each choice of parent factor, we simulate $1\,000$ different outbreaks and show the mean infections prevented per isolation (and error on the mean) in these simulations as compared to simulations on the same networks, with no contact tracing and the same set of initially infected nodes. For the {\BA} network, the contact tracing efficiency drops monotonically as forward contact tracing becomes a more important part of the contact tracing strategy. The opposite is true for the {\ER} network.
    }
    \label{fig:child_vs_parent}
\end{figure}

One might wonder whether releasing quarantined susceptible nodes changes the results presented in Figure~\ref{fig:flat_vs_empirical}A. 
Figure~\ref{fig:child_vs_parent} demonstrates that this is not the case; releasing quarantined susceptible nodes does not change that backward tracing is more efficient than forward tracing, at least when simulating the constant infectiousness disease model on a {\BA} network with mean degree $4$. In Figure~\ref{fig:child_vs_parent}, we plot the estimated number of infections prevented per isolation for different choices of contact tracing strategy. On the horizontal axis, we gradually decrease the weight that a direct source of infections is given in the contact list; from $w_{\rm parent}=1$ (and $w_{\rm non-parent}=1-w_{\rm parent}=0$) at the leftmost point, to $w_{\rm parent}=0$ at the rightmost point. The background color changes from magenta to green as the contact tracing scheme changes from pure backward contact tracing to pure forward contact tracing tracing as $w_{\rm parent}$ decreases from $1$ at the leftmost point of the plot to $0$ at the rightmost point (magenta and green being the face colors of the histograms in Fig.~\ref{fig:flat_vs_empirical}). For the Barabasi-Albert network, the efficiency of contact tracing decreases monotonically as parent nodes are given less weight. Changing the network to a fully connected {\ER} network with $245\, 046$ nodes and a mean degree of $4.08$ makes contact tracing efficiency monotonically increase as parent nodes are given less weight in the contact list.

\subsection{Replicating high estimates of contact-tracing efficiency}
In the previous sections, we showed that Kojaku \textit{et al.}'s~\cite{kojaku_effectiveness_2021} estimates of contact tracing efficiency are an order of magnitude higher than ours. Our epidemic model is different from Kojaku \textit{et al.}'s~\cite{kojaku_effectiveness_2021} in a number of ways. For example, time progresses in discrete time steps in our model, and is continuous in Kojaku \textit{et al.}'s~\cite{kojaku_effectiveness_2021}. Such differences can impact results significantly (see reference~\cite{juul_comparing_2021} for a demonstration of how offspring distributions are different for the Susceptible-Infected-Recovered model and the Independent Cascade model). It is therefore natural to wonder whether the large difference in estimates of contact tracing efficiency could be due to such model differences, and not the reasons outlined above. To test this possibility, we simulated contact tracing like Kojaku \textit{et al.}~\cite{kojaku_effectiveness_2021} did, by first running the epidemic to completion (using our constant infectiousness model), and then implementing mitigation measures on the resulting transmission trees. We obtained $1\,000$ different simulated transmission trees and implemented simulated contact tracing $10$ times on each. This gave us $10\,000$ estimates of contact tracing efficiency using this sequential method.

\begin{figure}
    \centering
    \includegraphics[width=0.96\linewidth]{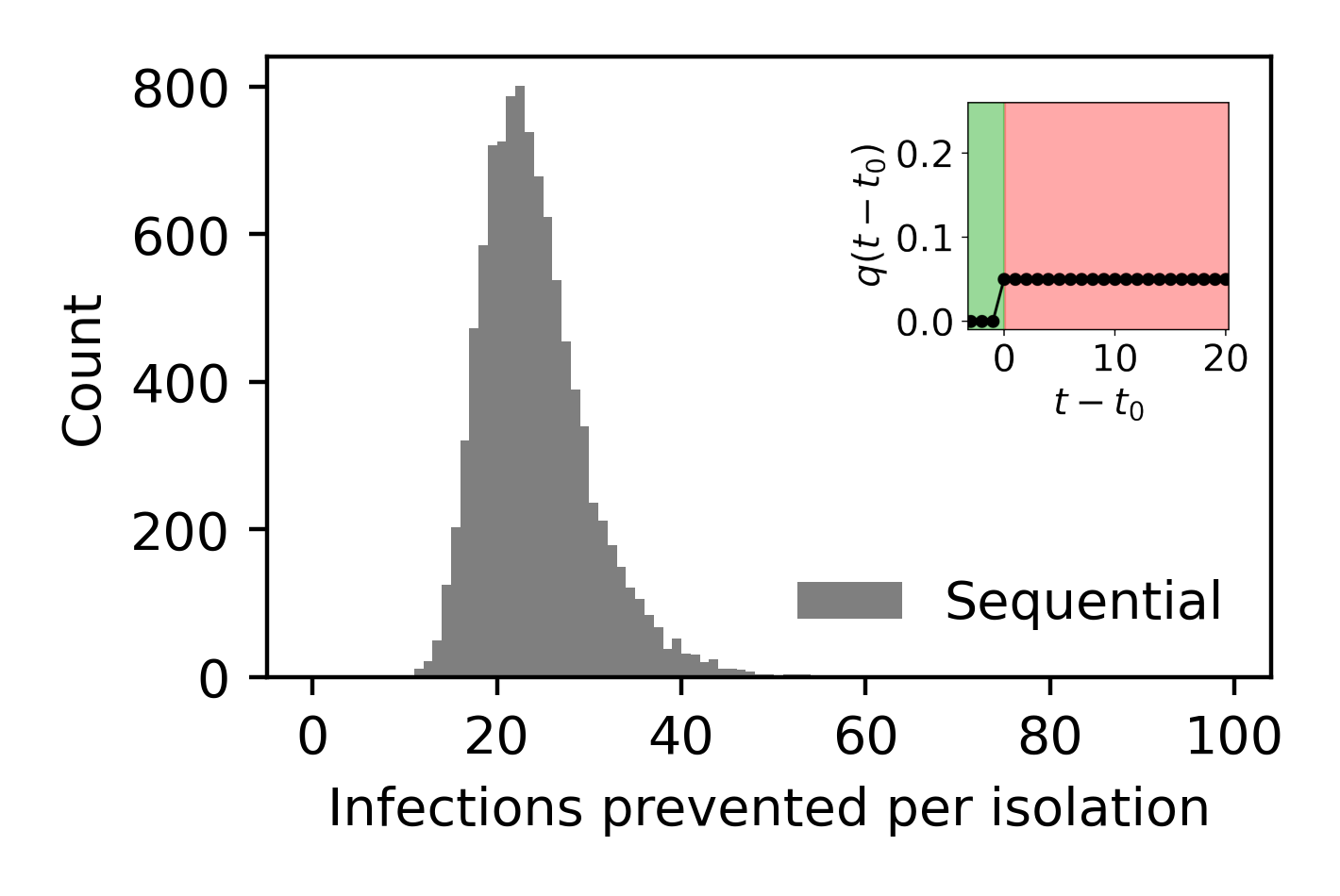}
    \caption{\textbf{Estimates of infections prevented per isolation when disease spread and mitigation measures are simulated sequentially rather than in parallel.} We simulated an infectious disease spreading to completion $1\,000$ times, obtaining $1\,000$ transmission trees in the process. On each tree, we simulated case isolation and contact tracing $10$ different times. Each time, we estimated the number of infections prevented per isolation. The figure shows the distribution of the $10\,000$ estimates. The mean value is $24.14\pm0.06$, a value similar to that obtained by Kojaku \textit{et al.}, but an order magnitude larger than the more realistic estimates obtained when simulating disease spread and mitigation measures in parallel.
    }
    \label{fig:Kojaku_estimate}
\end{figure}

Figure~\ref{fig:Kojaku_estimate} plots the results in a histogram. The distribution has a thick right tail and the resulting mean estimate of contact tracing efficiency was $24.14\pm0.06$ infections prevented per isolation. This is of the same order of magnitude as Kojaku \textit{et al.}'s estimate, and thus strongly supports the claim that the sequential nature of how Kojaku \textit{et al.}~\cite{kojaku_effectiveness_2021} simulate contact tracing causes their estimate of contact tracing efficiency to be much too large.

\section{Discussion}
Contact tracing is a central component in mitigation strategies for many infectious diseases. A better understanding of how different contact tracing strategies compare in terms of efficiency could translate into saved lives and decreased economic losses when faced with an epidemic. In this paper, we investigated recent influential claims that backward contact tracing is ``profoundly more effective'' than forward contact tracing, that contact tracing effectiveness ``hinges on reaching the `source' of infection'' and that contact tracing beats case isolation in terms of cost efficiency~\cite{kojaku_effectiveness_2021}.

By correcting shortcomings in how disease spread and mitigation measures were simulated in the  study~\cite{kojaku_effectiveness_2021}, we showed that the above-mentioned findings do not hold up. 
We conclude that contact tracing is not necessarily more cost efficient than case isolation, and that whether backward tracing or forward tracing is superior depends on the disease in question. For COVID-19-like disease dynamics, we found that backward tracing could actually be the less efficient strategy. Even so, backward tracing could still be valuable as a means to uncover new branches of the transmission tree that could then be forward traced. How efficient this strategy would be remains an open question.

We have simulated an idealized model for disease spread and mitigation on static {\BA} networks. The idealized nature of both the choices of models and networks means that simulations could be made more realistic by choosing empirically observed network structures, rather than the synthetic {\BA} networks, or by choosing disease models and parameters more carefully to match those of the COVID-19 pandemic. We stress, however, that the point of this work is not to present a realistic simulation of disease spread and disease mitigation. Instead, our main contribution is to show that recent conclusions about the superiority of backward tracing to forward tracing and case isolation are flawed and to demonstrate that the flaw stems not from unrealistic models of disease spread or mitigation measures, but from the sequential nature of Kojaku \textit{et al.}'s~\cite{kojaku_effectiveness_2021} simulations. This sequential simulation of disease spread and mitigation efforts cause both the quantitative estimates of backward tracing efficiency and the qualitative results comparing the efficiency of different mitigation strategies to be unreliable.

There are many interesting directions for future research on the efficiency of contact tracing strategies. As already mentioned, one direction is to investigate mixed strategies of strategically combined backward and forward contact tracing. Using a branching process framework Endo \textit{et al.}~\cite{endo_implication_2021} recently showed that a combination of backward and forward contact tracing can curb the spread of diseases with overdispersion better than forward tracing alone. Among other things, their analysis did not investigate the effect of time-varying infectiousness on contact tracing effectiveness, and many interesting questions remain.   How should resources for backward and forward contact tracing be allocated when faced with a specific disease? How many forward tracings should be carried out for each person that was successfully backward traced? Under what circumstances is backward contact tracing so inefficient that it should not be prioritized explicitly? Questions like these abound. 

Another interesting direction would be to estimate contact tracing efficiency at different stages of an epidemic. Contact tracing is of course most efficient in the very beginning of an epidemic, when there is still hope that the spreading can be stopped altogether. But how fast does the expected number of prevented cases per isolation drop as the epidemic infects more and more people? Is there a time when authorities should prioritize carrying out more contact tracing and another where resources are better spent increasing population screening capabilities? Answers to such questions would be useful for decision makers next time the world is faced with a pandemic.

\section{Code availability}
The code and data necessary to reproduce our results are available at
\texttt{https://github.com/jonassjuul/backward-tracing}.
\bibliographystyle{unsrt}
\bibliography{backward_bib.bib}

\end{document}